\documentclass[aps,pra,twocolumn,showpacs,floatfix,superscriptaddress, graphics, longbibliography]{revtex4-1}

\usepackage{amsmath,amssymb,graphicx,color,braket,hyphenat,makeidx,xcolor,dsfont,subfigure}
\usepackage[ocgcolorlinks,colorlinks=true,linkcolor=blue,citecolor=red,linktocpage=true]{hyperref}

\newcommand{\blue}{\color{\blue}}

\newcommand{\tr}{\mathrm{Tr}}
\newcommand{\R}{\mathcal{R}}

\newcommand{\gammat}{\gamma_{\{0,t\}}}
\newcommand{\gammatau}{\gamma_{\{0,\tau\}}}
\def\id {{\mathds 1}}

\begin{document}

\title{Quantum Martingale Theory and Entropy Production}

\author{Gonzalo Manzano}
\email{gmanzano@ictp.it}
\affiliation{International Center for Theoretical Physics ICTP, Strada Costiera 11, I-34151, Trieste, Italy}
\affiliation{Scuola Normale Superiore, Piazza dei Cavalieri 7, I-56126, Pisa, Italy}

\author{Rosario Fazio}
\affiliation{International Center for Theoretical Physics ICTP, Strada Costiera 11, I-34151, Trieste, Italy}
\affiliation{NEST, Scuola Normale Superiore and Instituto Nanoscienze-CNR, I-56126 Pisa, Italy}

\author{\'Edgar Rold\'an}
\email{edgar@ictp.it}
\affiliation{International Center for Theoretical Physics ICTP, Strada Costiera 11, I-34151, Trieste, Italy}

\begin{abstract}
We employ martingale theory to describe fluctuations of entropy production for open quantum systems in nonequilbrium steady states. Using the formalism of quantum jump trajectories, we identify a decomposition of entropy production into an exponential martingale and a purely quantum term, both obeying integral fluctuation theorems. An important consequence of this approach is the derivation of a set of genuine  universal results for stopping-time and infimum  statistics of stochastic entropy production. Finally we complement the general formalism with numerical simulations of a qubit system.
\end{abstract}


\maketitle 

The development of stochastic thermodynamics in the last decades allowed the description of work, heat and entropy production at the level of single trajectories in nonequilibrium processes~\cite{seifert, lebowitz}. 
This framework has successfully provided several genuine insights on the second law, such as the discovery of  universal relations constraining the statistics of fluctuating thermodynamic quantities, 
usually known as fluctuation theorems~\cite{sekimoto,SeifertREV, JarzynskiREV}. The fundamental interest in refining our understanding of irreversibility and their microscopic imprints has been brought to 
its ultimate consequences by extending stochastic thermodynamics to the quantum realm~\cite{quantum}, where fluctuation theorems have been derived~\cite{fluctuations1, fluctuations2, deffner, tasaki, ueda, manzano, bartolotta}, and experimentally tested in the last years~\cite{experimentalFT1, experimentalFT2}.

When information about entropy production in single trajectories of a process is available, a natural question to ask is until what extend this information can be useful. For instance whether or not it is possible to implement strategies 
leading to a reduction in entropy which might be eventually used as a fuel, like in the celebrated Maxwell's demon~\cite{demons1,demons2}. In the same context, one may ask whether the second law of thermodynamics will manifest as fundamental constraints limiting such strategies. 
A powerful method to handle these general questions, is to employ a set of particularly interesting  stochastic processes, namely Martingales~\cite{martingale0}. Martingales are well known in mathematics~\cite{martingales} and quantitative finance as models of fair financial markets~\cite{economy}. However, martingale theory has been only little exploited until now both in stochastic thermodynamics~\cite{neri,raphaelshamik,haoge,ventejou,singh, chetrite,pigolotti} and quantum physics~\cite{bauer,adler}.

Applying concepts of martingale theory in stochastic thermodynamics, it has been shown that the exponential of minus the  entropy production $\Delta S_\mathrm{tot}(t)$ associated with classical trajectories $\gammat$  $-$single paths in phase space$-$  in generic non-equilibrium steady-state conditions is an \emph{exponential martingale}, i.e. $\langle e^{-\Delta S_\mathrm{tot}(t)}\, |\,\gammatau \rangle = e^{-\Delta S_\mathrm{tot}(\tau)}$, for any $t\geq \tau\geq 0$, where  $\langle X(t) | \gammatau \rangle$ denotes the conditional expectation of a functional $X(t)$ given $\gammatau$~\cite{martingales}. 
It has been shown that the martingality of $e^{-\Delta S_\mathrm{tot}(t)}$ implies a series of universal equalities and inequalities concerning  the statistics of infima and stopping times (e.g. first-passage, escape times, etc.) of entropy production~\cite{neri,chetrite}. 

\begin{figure}
 \includegraphics[width= 1.0 \linewidth]{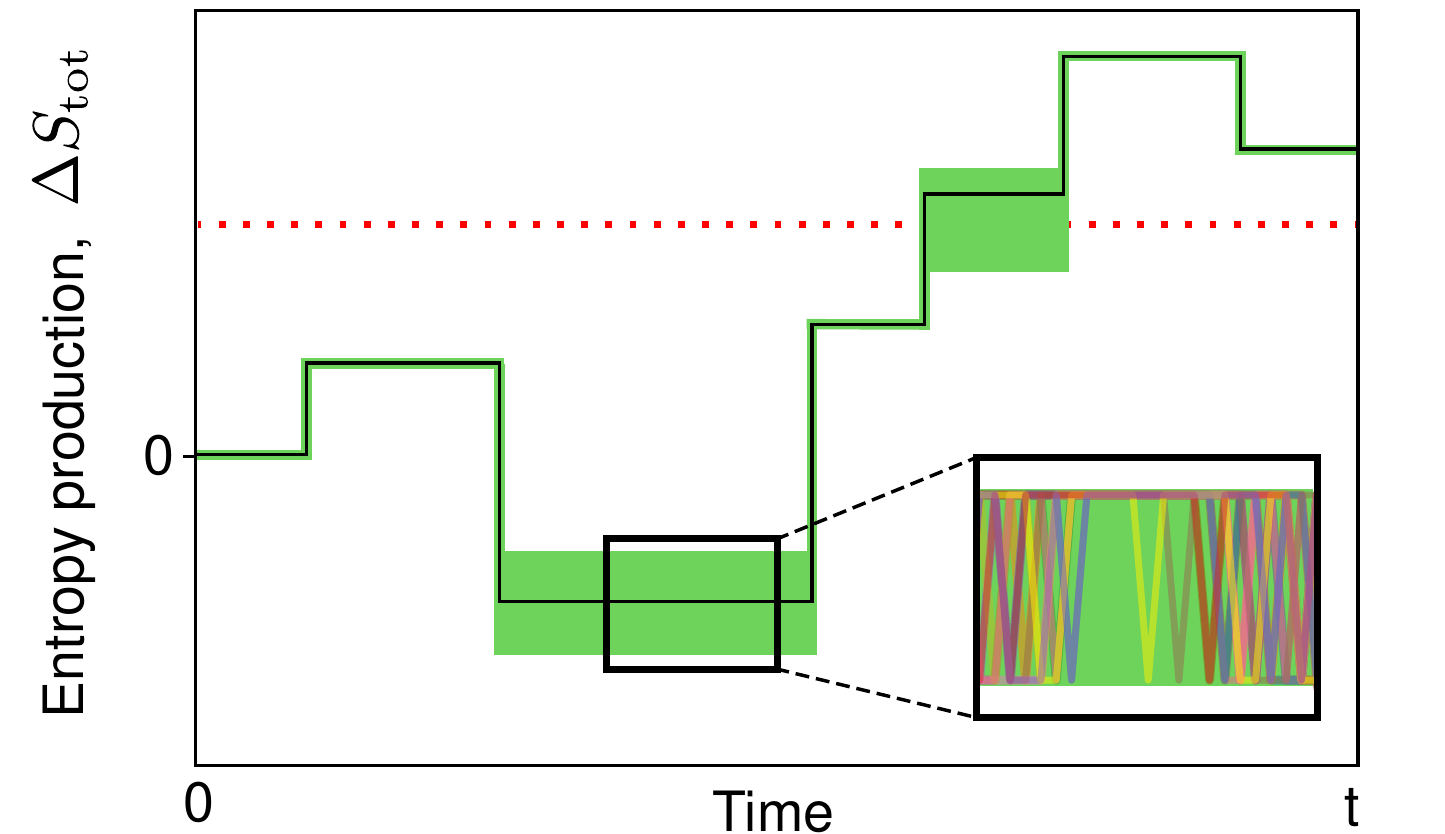}
 \caption{Example traces of stochastic entropy production in a nonequilibrium stationary process as a function of time. We show the entropy production $\Delta S_\mathrm{tot}$ associated with classical (black thin line) and quantum (thick green line) trajectories of duration $t$. 
 When the system state  becomes a superposition, $\Delta S_\mathrm{tot}$ is not uniquely defined in the quantum case (thick green segments)  but only at the final time of the evolution $t$, when a direct measurement on the system is performed. The inset shows different values that  $\Delta S_{\rm tot}$  would take if system measurements were performed at intermediate times,   
 for a given record of measurements in the environment. How can then one determine stopping times e.g. when does entropy production reach a threshold (red dotted line)  for the first time in an open  quantum  system?} \label{fig1}
 \end{figure}

In this Letter, we generalize the martingale theory of entropy production ---so far only developed for classical systems--- to the context of quantum thermodynamics, using the formalism of quantum jump trajectories~\cite{milburn, trajectories}. 
Genuine quantum effects (such as coherence~\cite{coherence1,coherence2,coherence3} and quantum correlations~\cite{correlations1,correlations2,correlations3}) introduce new qualitative features that radically modify the framework. Indeed, dealing with martingales and 
stopping times requires conditioning on past events, which entails several difficulties in the quantum realm. 
Since evaluating stochastic entropy production along quantum trajectories $\gamma_{\{0,t\}}$ requires a two-point measurement protocol using direct measurements on the system~\cite{deffner, horowitz, horowitz2, manzanoPRE, manzano, auffeves, gheradini, landi}, the development of a suitable conditioning not disturbing the dynamics becomes challenging.
See Fig.~\ref{fig1} for an illustration.

In the following, we introduce a new auxiliary entropy production $\Delta S_{\rm mar}(t)$ and show that $e^{-\Delta S_{\rm mar}(t)}$ is a martingale along quantum trajectories [Eq.~\eqref{eq:quantummartingale} below], while in general $e^{-\Delta S_{\rm tot}(t)}$ is not. 
We use this finding to derive several universal relations valid for all nonequilibrium steady states providing new insights on the role of uncertainty and coherence in the second law. Our key results are established in Eqs.~(\ref{eq:quantummartingale}-\ref{eq:ILQ}). They comprise: (i) a genuine quantum-classical decomposition of stochastic entropy production as the sum of two quantities, both fulfilling  integral fluctuation theorems [Eqs.~\eqref{eq:decomp} and~\eqref{eq:3ft}], entailing a tight lower bound for the average entropy production [Eq.~\eqref{eq:EPbound}]; (ii)  fluctuation theorems and inequalities for stopping-time statistics of entropy production [Eqs.~(\ref{eq:Doob}-\ref{eq:Doob2})]; and  (iii) inequalities  for the extreme-value statistics of entropy production that  generalize to the quantum realm the results derived in Refs.~\cite{neri,raphaelshamik} [Eqs.~(\ref{eq:infima}-\ref{eq:ILQ})].

{\it Entropy production in quantum trajectories ---} One of the most successful approaches describing the stochastic thermodynamics of open  quantum systems is the formalism of quantum jump trajectories~\cite{hekking,campisi,romito, gong, horowitz, horowitz2, manzanoPRE, auffeves, gheradini, landi, elouard, naghiloo, circuits,demon}. This formalism describes the stochastic evolution of the pure state of the system $\ket{\psi (t)}$, conditioned on measurements  obtained from the continuous monitoring of the environment~\cite{milburn,trajectories}. 

Within this approach, the system evolution is described as a smooth evolution intersected by quantum jumps in the state of the system 
occurring at random times. These jumps correspond to the detection of different type of events in the environment (e.g. the emission or absorption of energy quanta from different thermal reservoirs) leading to a measurement record $\R_{0}^{t} = \{(k_1, t_1), ..., (k_J, t_J)\}$, 
where $(k_j, t_j)$ means that a jump of type $k_j$ occurred at time $t_j$, $j= 1, ... J$, and $0\leq t_1\leq \dots \leq t_J\leq t$. The evolution is described by the stochastic Schr\"odinger equation:
\begin{align} \label{eq:stochastic}
 \text{d} \ket{\psi (t)}&=\text{d}t \left( -\frac{i}{\hbar} H  + \sum_k\frac{\langle L_k^\dagger  L_k  \rangle_{\psi (t)} - L_k^\dagger  L_k}{2}\right)\!\ket{\psi (t)} \nonumber \\
 &+ \sum_k \text{d}N_k(t) \left( \frac{L_k }{\sqrt{\langle L_k^\dagger L_k \rangle_{\psi(t)}}} - \id \right)\!\ket{\psi(t)},
\end{align}
where $H$ is an hermitian operator (usually the system Hamiltonian),  $L_k$ for $k = 1 ... K$ are the Lindblad (jump) operators, and here and in the following we  denote as $\langle A \rangle_{\psi (t)} \equiv \langle \psi(t)| \,A \,|\psi(t) \rangle$ quantum-mechanical expectation values, and $\id$  the identity matrix. 
The random variables $\text{d}N_k(t)$  are Poisson increments associated to the number of jumps $N_k(t)$ of type $k$ detected up to time $t$ in the process, leading to the record $\R$. The $\text{d}N_k(t)$ take most of the time the value $0$, becoming $1$ only at times $t_j$ when a jump of type $k_j$  is detected in the environment. 
When  averaging over measurement outcomes, the evolution reduces to a Markovian process ruled by a Lindblad master equation~\cite{Lindblad, rivas}. 

From now on, we consider nonequilibrium steady states.  Here, the initial state of the trajectories is sampled from the spectral decomposition of the steady state of the master equation, $\pi = \sum_n \pi_n \ket{\pi_n}\!\bra{\pi_n}$. We also require a two-measurement protocol, where projective measurements in the $\pi$-eigenbasis are executed at the beginning ($t=0$) and at the end ($t=\tau$) of any single trajectory. For this setup, a probability $P(\gamma_{\{0,\tau\}})$ can be associated to any trajectory $\gamma_{\{0,\tau\}} = \{n(0); \R_0^\tau;n(\tau) \}$, where $n(0)$ and $n(\tau)$ are the outcomes of the first and final measurements on the system. The entropy production (for $k_B=1$) associated with the trajectory $\gammatau$ is defined as the functional:
\begin{equation}\label{eq:ep}
 \Delta S_\mathrm{tot}(\tau) \equiv \ln \frac{P (\gammatau)}{ \tilde{P}(\tilde{\gamma}_{\{0,\tau\}})} =   \ln  \left[\frac{\pi_{n(0)}}{\pi_{n(\tau)}}\right]   +  \sum_{j = 1}^J \Delta S_{\rm env}^{k_j},
\end{equation}
where $ \tilde{P}(\tilde{\gamma}_{\{0,\tau\}})$ is the probability of the time-reversed trajectory $\tilde{\gamma}_{\{0,\tau\}} = \{ n(\tau); \tilde{R}_\tau^0; n (0) \}$, with $\tilde{\R}_\tau^0 = \{ (k_J,t_J) , ..., (k_1,t_1)\}$ the time-reversed sequence of jumps, 
occurring in the time-reversed (or backward) process~\cite{manzano, horowitz, horowitz2}.
In Eq.~\eqref{eq:ep}, the first term in the r.h.s. is the system entropy change along the trajectory~\cite{seifert}, and  
$\Delta S_{\rm env}^{k_j}$ as the environmental entropy change due to the jump $k_j$, which in most cases of physical interest obey the local detailed balance condition for pairs of operators $L_k^{~} =  L_{k^\prime}^\dagger e^{\Delta S_{\rm env}^{k}/2}$~\cite{horowitz2}.
 The averages of both terms yield the von Neumann entropy changes of system and environment, respectively~\cite{manzano}.
The stochastic entropy production given by Eq.~\eqref{eq:ep} obeys the integral fluctuation theorem $\langle e^{-\Delta S_\mathrm{tot}(\tau)} \rangle = 1$, which  leads to the second law  inequality $\langle \Delta S_\mathrm{tot} (\tau) \rangle \geq 0$.
The classical limit is recovered when the stochastic wavefunction $\ket{\psi(t)}$ obtained from~\eqref{eq:stochastic} is an eigenstate of $\pi$ at any time of the dynamical evolution~\cite{climit}. 

{\it Quantum martingale theory ---} The main ingredient for the development of our quantum martingale theory is  the definition of conditional averages over trajectories with common history up to a certain time $\tau\leq t$. To this end, one needs to define both $\gammatau$ and $\gammat$. Here, unlike for classical trajectories, $\gammatau\nsubseteq\gammat$ because $\gammatau$ includes a measurement at time $\tau$ while $\gammat$ does not, and therefore entropy production for trajectory $\gammat$ is not well-defined at time $\tau$ (see Fig.~\ref{fig1}). Thus, we define the conditional average of a generic stochastic process $X(t)$ defined along a trajectory $\gammat$ as $\langle X(t) | \gamma_{[0,\tau]} \rangle = \sum_{n(t) , \R_\tau^t} X(t) P(\gammat | \gamma_{[0,\tau]})$, where we condition with respect to the {\it ensemble} of trajectories $\gamma_{[0, \tau]} \equiv \bigcup_{s=0}^\tau \gamma_{\{0,s\}}$ that includes all the outcomes of  trajectories eventually stopped (i.e. measured) at all intermediate times in the interval $[0,\tau]$.   Note that in the classical limit the ensemble $\gamma_{[0,\tau]}=\gammatau$ just contains one trajectory. Furthermore, since $P(\gammat|\gamma_{[0,\tau]})=P(\gammat|\gammatau)$ (see Supplemnental Material~\cite{supplemental}), then $\langle X(t) | \gamma_{[0,\tau]} \rangle = \langle X(t) | \gammatau \rangle $.

We now discuss quantum martingales in relation with  stochastic entropy production given by Eq.~\eqref{eq:ep}. Notably, unlike for  classical systems~\cite{neri,raphaelshamik}, the process $e^{-\Delta S_\mathrm{tot} (t)}$   is not  a martingale in this context because $\langle e^{-\Delta S_\mathrm{tot} (t)} | \gamma_{[0,\tau]}\rangle =e^{-\Delta S_\mathrm{tot} (\tau)+\Delta S_{\rm unc}(\tau) } $ for all $t\geq \tau\geq 0$, with
\begin{equation}\label{eq:spsi}
\Delta S_{\rm unc}(t) \equiv -  \ln \left[ \frac{\pi_{n(t)}}{\langle \pi \rangle_{\psi(t)}} \right],
\end{equation}
see~\cite{supplemental} for a detailed proof. The origin of $\Delta S_{\rm unc}$ can be traced back to the quantum uncertainty in the evolution. It measures how informative (surprising) is the occurrence of outcome $n(t)$ with respect to the average result when measuring the stochastic wavefunction $\ket{\psi(t)}$ at that time, that is, $\langle \pi \rangle_{\psi(t)} = \sum_i \pi_i |\langle \pi_i |\psi(t) \rangle|^2$. 
The quantity $\langle \pi \rangle_{\psi(t)}$ is  the squared Ulhman's Fidelity between states $\pi$ and $\ket{\psi(t)}$, which quantifies the distinguishably of these two states. In other words, $\Delta S_{\rm unc}(t)$ measures how much information we gain knowing the outcome $n(t)$ of the measurement at time $t$ with respect to knowing only $\ket{\psi(t)}$. 
Importantly, the ``uncertainty'' entropy production $\Delta S_{\rm unc}(t)$ satisfies  the property $\langle e^{-\Delta S_{\rm unc}(t)} | \gamma_{[0,\tau]} \rangle = 1$ for any $\tau \leq t$~\cite{supplemental}, and it is bounded at all times  by $\pi_\mathrm{min}/\pi_\mathrm{max} \leq e^{-\Delta S_{\rm unc}(t)} \leq \pi_\mathrm{max}/\pi_\mathrm{min}$, where $\pi_\mathrm{min} = \min_i \pi_i$ and $\pi_\mathrm{max} = \max_i \pi_i$ are the minimum and maximum eigenvalues of $\pi$.
In the classical limit 
$\langle \pi \rangle_{\psi(t)} = \pi_{n(t)}$ in Eq.~\eqref{eq:spsi}, leading to $\Delta S_{\rm unc}(t) = 0$ at any time $t$, and we recover the classical result.

We now define the auxiliary ("martingale") entropy production  as $\Delta S_\mathrm{mar}(t) \equiv  \Delta S_\mathrm{tot}(t) - \Delta S_{\rm unc}(t)$, which using Eqs.~\eqref{eq:ep} and~\eqref{eq:spsi} gives
\begin{equation}\label{eq:aux}
 \Delta S_\mathrm{mar}(t) =   \ln  \left[\frac{\pi_{n(0)}}{\langle \pi \rangle_{\psi(t)}}\right]   +  \sum_{j = 1}^J \Delta S_{\rm env}^{k_j},
\end{equation}
to be compared with Eq.~\eqref{eq:ep}. Note that $ \Delta S_\mathrm{mar}(t)$ results from replacing  $\pi_{n(t)}$   by $\langle \pi \rangle_{\psi(t)}$ in the boundary term in~\eqref{eq:ep}, therefore avoiding the need of information from measurements at time $t$. 
We prove that $\Delta S_{\rm mar}(t)$ is an exponential martingale:
\begin{equation} \label{eq:quantummartingale}
 \langle e^{-\Delta S_\mathrm{mar}(t) } |\, \gamma_{[0,\tau]} \, \rangle = e^{-\Delta S_\mathrm{mar}(\tau)}\quad,
\end{equation}
which holds for any $t\geq \tau\geq 0$~\cite{supplemental}. Recall that the average in~\eqref{eq:quantummartingale} is conditioned over the ensemble of trajectories $ \gamma_{[0,\tau]}$ containing  all possible measurement outcomes  in both system and environment  at times smaller than $\tau$. However, since $\langle X(t) | \gamma_{[0,\tau]} \rangle = \langle X(t) | \gammatau \rangle$ for any functional $X(t)$ of $\gammat$, one also  has a martingale condition with respect to single quantum trajectories, $ \langle e^{-\Delta S_\mathrm{mar}(t) } |\, \gamma_{\{0,\tau\}} \, \rangle = e^{-\Delta S_\mathrm{mar}(\tau)}$. 

\begin{figure*}[t]
 \includegraphics[width= 1.0 \linewidth]{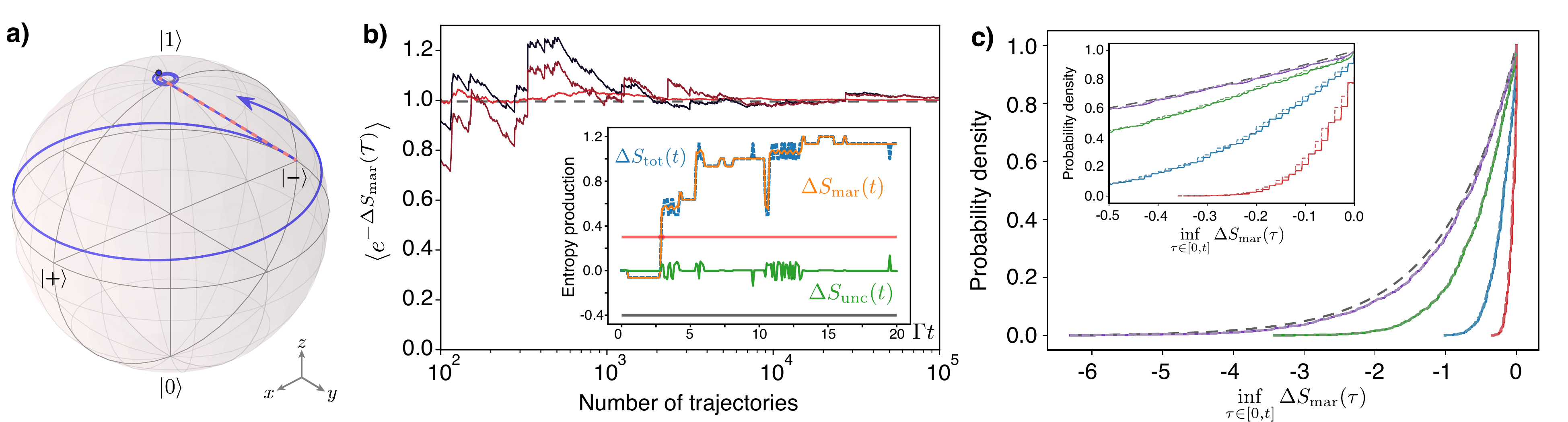}
 \caption{(a) Sample trajectory of $\ket{\psi (t)}$ represented in the qubit's Bloch sphere. The smooth evolution starting from an eigenstate (blue circle) is interrupted by a jump $L_{-}$ (dotted line) which abruptly collapses the qubit state to $\ket{-} = [\ket{0} - \ket{1}]/\sqrt{2}$ in the $x$-axis, after which the evolution continues up to the point marked by the arrow.  
 (b) Convergence of the stopping-time fluctuation theorem for $\Delta S_{\rm mar}$  as a function of the number of trajectories for three different stopping rules: stopping at a fixed time $t_\mathrm{f}  \equiv 20 \Gamma^{-1}$  (black  line); $\min (\mathcal{T}_1,t_{\rm f})$ with $\mathcal{T}_1$ the first-passage time of $\Delta S_{\rm mar}$ to reach a positive threshold located at $0.3$ (brown  line); $\min (\mathcal{T}_2,t_{\rm f})$ with $\mathcal{T}_2$ the unconditional first-passage time of $\Delta S_{\rm mar}$ to reach either the thresholds  $0.3$ or $-0.4$ (red  line). The dashed grey line in $1$ is a guide to the eye.  Inset:  example traces of stochastic entropy production (blue dashed line) and its decomposition as the sum of $\Delta S_{\rm mar}$ (orange solid line) plus $\Delta S_{\rm unc}$ (green solid line). The horizontal thick lines are the  two absorbing boundaries  used to compute stopping times. Parameters of the simulation: $ \hbar \omega = 1$, $\beta = 0.2$, $\eta = 0.5$, $\Gamma\equiv\gamma_- - \gamma_+ = \gamma_\downarrow - \gamma_\uparrow = 0.01$. 
 (c) Empirical cumulative distributions of the finite-time minimum of $\Delta S_{\rm mar}$ (solid) and $\Delta S_{\rm tot}$ (dashed) for different values of the observation time: $\Gamma t=10$ (red), $\Gamma t=10^2$ (blue), $\Gamma t=10^3$ (green) and $\Gamma t=10^4$ (purple). The grey dashed line is the exponential in the right-hand side of Eq.~\eqref{eq:infima}. The inset shows a zoomed view of the distribution for small values of the infima. Parameters or the simulation: $ \hbar \omega = 1$, $\beta = \eta = 0.04$, $\Gamma= 0.01$.} \label{fig2}
 \end{figure*}

{\it Martingale fluctuation theorems and  second law---} The first consequence of  Eq.~\eqref{eq:quantummartingale}, together with the properties of $\Delta S_{\rm unc}(t)$ in Eq.~\eqref{eq:spsi} is  the following decomposition of stochastic entropy production
\begin{equation}\label{eq:decomp}
\Delta S_{\rm tot}(t) = \Delta S_{\rm unc} (t) + \Delta S_{\rm mar}(t),
\end{equation}
with both summands in~\eqref{eq:decomp} fulfilling an integral fluctuation theorem:
\begin{equation}\label{eq:3ft}
\langle e^{-\Delta S_{\rm unc}(t)}\rangle =1 ,\quad \langle e^{-\Delta S_{\rm mar}(t)}\rangle=1.
\end{equation}
Notably, this decomposition has the same structure as the Oono-Paniconi~\cite{oono} (and adiabatic/non-adiabatic~\cite{ev}) decomposition of entropy production in nonequilibrium systems~\cite{harris}. From Jensen's inequality $\langle e^{x} \rangle\geq e^{\langle x \rangle} $ and Eq.~\eqref{eq:3ft},   we have  both $\langle \Delta S_{\rm unc}(t)\rangle \geq 0$ and  $\langle \Delta S_{\rm mar}(t)\rangle \geq 0$. Moreover, using~\eqref{eq:decomp} and the second law, we obtain the following bound for the average  entropy production
\begin{equation} \label{eq:EPbound}
\langle \Delta S_\mathrm{tot}(t) \rangle \geq  \langle \Delta S_{\rm mar} (t) \rangle.
\end{equation}
We remark that the bound~\eqref{eq:EPbound} allows to  estimate  the average entropy production without the need of any measurement on the system at intermediate times.
This inequality provides a  tight bound because $\Delta S_{\rm unc}(t)$ is bounded and thus not extensive in time~\cite{VN}.  
In the classical limit and when approaching equilibrium conditions $\Delta S_{\rm unc}(t)  = 0$, and~\eqref{eq:EPbound} becomes an equality. 

{\it Stopping-time statistics ---}  An important result in martingale theory is Doob's optional stopping theorem~\cite{doob} which concerns stopping-time statistics of martingales. A paradigmatic example of a stopping time $\mathcal{T}$ is the first time at which a  stochastic process $X(t)$ reaches a  subset $\mathcal{X}$ of the state space.  Importantly, a stopping time  $\mathcal{T}$ is a random variable whose value can be determined solely by looking at the past history of the process $X_{[0,\mathcal{T}]}$.
Applying Doob's optional stopping theorem to the martingale $e^{-\Delta S_{\rm mar}(t)}$ one finds 
\begin{equation}\label{eq:Doob}
\langle e^{-\Delta S_\mathrm{mar} (\mathcal{T}) } \rangle= 1, 
\end{equation}
i.e. its  average over the stopping times $\mathcal{T}$ equals its average value at the initial time $t=0$~\cite{supplemental}. Again, using Jensen's inequality and Eq.~\eqref{eq:Doob} we obtain a second-law-like inequality at stopping times $\langle\Delta S_\mathrm{mar} (\mathcal{T})  \rangle\geq 0$ which implies 
\begin{equation}\label{eq:Doob2}
\langle  \Delta S_\mathrm{tot} (\mathcal{T}) \rangle \geq  \langle \Delta S_\mathrm{\rm unc} (\mathcal{T}) \rangle.
\end{equation}
Note that here $\mathcal{T}$ are stopping times defined in terms of $\gamma_{[0,\mathcal{T}]}$ which are well defined when the stopping condition uses the auxiliary process $\Delta S_{\rm mar}(t)$. 
In the classical limit $\Delta S_{\rm unc}(t)=0$, Eq.~\eqref{eq:Doob} reduces to the integral fluctuation theorem for entropy production at stopping times $\langle e^{-\Delta S_\mathrm{tot} (\mathcal{T}) } \rangle= 1$, and Eq.~\eqref{eq:Doob2} to the second law $\langle  \Delta S_\mathrm{tot} (\mathcal{T}) \rangle\geq 0$. We remark that $\langle \Delta S_{\rm unc}(\mathcal{T})\rangle$ can in principle be either positive or negative. Therefore, Eq.~\eqref{eq:Doob2} does not exclude the possibility  that the average entropy production at stopping times may be \textit{negative} for  particular choices of stopping times.

{\it Extreme-value statistics ---}  Very recently, universal statistics of infima of stochastic entropy production have been unveiled~\cite{neri} using Doob's maximal inequality~\cite{doob} which bounds the probability of the supremum  of a positive martingale process $M(t)$ as $\mathsf{Pr}(\mathrm{sup}_{\tau \in [0,t]} M(\tau) \geq \lambda) \leq  \langle M(t) \rangle/\lambda$ for $\lambda\geq 0$.
Applying Doob's maximal inequality to the positive martingale $e^{-\Delta S_\mathrm{mar}(t)}$ we derive the following inequality for the probability that the finite-time infimum $\inf_{\tau\in[0,t]} \Delta S_\mathrm{mar}(\tau) \geq 0$ lies below a certain value
\begin{equation} \label{eq:infima}
\mathsf{Pr}\left(\inf_{\tau\in[0,t]} \Delta S_\mathrm{mar}(\tau) \leq -\xi \right) \leq e^{-\xi}, 
\end{equation}
where $\xi \geq 0$ and $t\geq 0$~\cite{supplemental}. Therefore, the probability to observe extreme reductions of the martingale entropy production along single trajectories of any duration $t$ is exponentially suppressed.  As a consequence, the average infimum of the martingale entropy production obeys $\left\langle\inf_{\tau\in[0,t]} \Delta S_\mathrm{mar}(\tau)  \right\rangle\geq  -1$, which generalizes the infimum law  for classical nonequilibrium stationary states~\cite{neri}. From Eq.~\eqref{eq:infima} and  the condition $\Delta S_{\rm unc}(t)\leq \ln [\pi_{\max}/\pi_{\min}]$ for all $t\geq 0$, we derive 
\begin{equation}\label{eq:ILQ}
\left\langle\inf_{\tau\in[0,t]} \Delta S_\mathrm{tot}(\tau)  \right\rangle\geq  -1 - \ln \left[\frac{\pi_{\max}}{\pi_{\min}}\right].
\end{equation}
Since $\ln[\pi_{\max}/\pi_{\min}]\geq 0$, Eq.~\eqref{eq:ILQ} provides   a  lower bound for the average entropy-production infimum that is below  the infimum law for classical systems  $\left\langle\inf_{\tau\in[0,t]} \Delta S_\mathrm{tot}(\tau)  \right\rangle\geq  -1$. 

{\it Quantum martingale theory at work ---} We conclude by illustrating our theory with a simple example, amenable of a direct experimental realisation. Our model consists of a single spin-$\frac{1}{2}$ particle (a  qubit) with Hamiltonian $H = \hbar \omega \sigma_z/2$ which is subjected to two different and orthogonal sources of noise.  The dynamical evolution is described by the stochastic Schr\"odinger equation~\eqref{eq:stochastic} where a monitoring process detects the jumps induced in the system by both noise sources. This can be visualized in the Bloch sphere where the stochastic wave function sits at all times during its evolution [see Fig.~\ref{fig2}(a)]. 
 The first source corresponds to thermal noise, which produces jumps in the $z-$direction of the particle as described by Lindblad operators $L_\downarrow = \sqrt{\gamma_\downarrow} \sigma^{-}$ and $L_\uparrow = \sqrt{\gamma_\uparrow} \sigma^{+}$, with  rates fulfilling local detailed balance $\gamma_\downarrow = \gamma_\uparrow e^{- \beta \hbar \omega}$, $\beta$ being the inverse temperature. The environmental entropy changes associated to these jumps are respectively $\Delta S^{\downarrow \uparrow}_\mathrm{env} = \pm \beta \omega$. Furthermore, a second source of noise generates jumps in the $x-$direction induced by the Lindblad operators $L_{-} = \sqrt{\gamma_{-}} (\sigma_{z} - i \sigma_{y})/2$ and $L_{+} = \sqrt{\gamma_{+}}(\sigma_{z} + i \sigma_{y})/2$, with $\gamma_{-} = \gamma_{+} e^{- \eta}$. Notice that here $\eta$ is a bias parameter that plays the  role of an inverse temperature in the $x-$direction. Analogously, the entropy changes associated to such jumps are  $\Delta S^{\mp}_\mathrm{env} = \pm \eta$.  The dynamical evolution in this setup is genuinely quantum. Since the two set of jumps occur in orthogonal directions,  the generation of superpositions of $\pi$-eigenstates along the quantum trajectories is guaranteed. The classical limit is  recovered for high temperatures ($\beta \rightarrow 0$) and large bias ($\eta \rightarrow 0$) where the steady state becomes the maximally-mixed state $\pi \rightarrow \id/2$, corresponding to the equilibrium distribution.

We performed numerical simulations of the qubit system using quantum-trajectory Montecarlo methods~\cite{qutip}. 
We find perfect agreement between our simulations and the stopping-times fluctuation theorem in Eq.~\eqref{eq:Doob} which is verified for three different stopping times $\mathcal{T}$ (Fig.~\ref{fig2}b). Interestingly, the convergence to the theoretical value is much faster for first-passage times over thresholds (see inset) than using a fixed (final) time as in the standard integral fluctuation theorem. This makes the stopping-time fluctuation theorem~\eqref{eq:Doob} more amenable for experimental tests.
We then evaluate statistics of the  finite-time minimum of $\Delta S_\mathrm{mar}$ and  $\Delta S_\mathrm{tot}$. The tails of the distribution of the finite-time minima of $\Delta S_\mathrm{mar}$ are exponentially suppressed in agreement with Eq.~\eqref{eq:infima} (Fig.~\ref{fig2}c). Moreover, the average minima of both  $\Delta S_\mathrm{mar}$ and $\Delta S_\mathrm{tot}$ lie  above $-1$ for this choice of parameters.

{\it Discussion ---} Our work shows that martingale theory can be generalised to quantum thermodynamics providing insights about entropy production  beyond fluctuation theorems. Here we provided new nonequilibrium universal relations along quantum trajectories [Eqs.~(\ref{eq:quantummartingale}-\ref{eq:ILQ})]. Our results may  be of particular importance in setups allowing environmental monitoring and feedback control~\cite{demon, naghiloo, Masuyama, carmichael, Dima, dicarlo, haroche, barreiro}, and for quantum thermal devices working in nonequilibrium steady-state conditions~\cite{Popescu, Kosloff, Giovannetti, Correa, Mitchison, Hofer, Scarani, Withney}. 
It would be also interesting to explore connections with path-integral approaches~\cite{funo}, one-shot quantum thermodynamics~\cite{Spekkens, Oppenheim, Halpern, Aberg}, and quantum information~\cite{Kempe,Cirac,Davide,Jordan, porras}. Finally, we remark that some of our results could be applied  to classical systems where knowledge of  system's state is incomplete e.g. under coarse-graining of hidden internal microstates~\cite{massicoarse,samurai}.

\begin{acknowledgments}
We  thank I. Neri, S. Singh, S. Suomela, I. Khaymovich, R. Chetrite, and S. Gupta for enlightening discussions. R.F. acknowledges support from SNS-WIS joint lab QUANTRA.
\end{acknowledgments}

\clearpage

\onecolumngrid

\widetext
\begin{center}
\textbf{\large Supplemental Material: Quantum Martingale Theory and Entropy Production}
\end{center}

\setcounter{equation}{0}
\setcounter{figure}{0}
\setcounter{table}{0}
\makeatletter
\renewcommand{\thesection}{S\arabic{section}}
\renewcommand{\theequation}{S\arabic{equation}}
\renewcommand{\thefigure}{S\arabic{figure}}
\renewcommand{\bibnumfmt}[1]{[S#1]}
\renewcommand{\citenumfont}[1]{S#1}

\newcommand{\Li}{\mathcal{L}}

\

The Supplemental Material contains detailed proofs of our main results as stated in Eqs.(5)-(12) of the main text. In section \ref{s1} we give further details on conditional averages in the context of quantum trajectories and provide the proofs of the main martingale properties of entropy production leading to Eqs.~(5)-(7). Section \ref{s2} is devoted to the proofs of the stopping-time fluctuation theorem in Eq.~(9) and the inequality~(10) for the average entropy production on stopping times. 
Finally, in section \ref{s3} we prove the inequality~(11) for the infima statistics of entropy production and the bound for the average finite-time infimum in Eq.~(12).

\section{Martingality proofs for stochastic entropy production} \label{s1}
In this section of the Supplemental Material we provide more details about the evaluation of conditional averages in the context of quantum trajectories, which are required to prove our three main results regarding the martingale property and stochastic entropy production [Eqs.~\eqref{proof1-6}, \eqref{proof2-3}, and \eqref{proof3-7} below]. 
We first prove that the total stochastic entropy production $\Delta S_\mathrm{tot}(t)$ in Eq.~(2) of the main text, is not an exponential martingale in general [Eq.~\eqref{proof1-6}]. 
Then we prove a generalized fluctuation theorem for the uncertainty entropy production $\Delta S_\mathrm{unc}(t)$ in Eq.~(3) [Eqs.~\eqref{proof2-3}]. Finally we also prove Eq.~(5), that is, the martingality of $e^{-\Delta S_\mathrm{mar}(t)}$, where $\Delta S_\mathrm{mar}(t)$ is the auxiliary (martingale) entropy 
production in Eq.~(4) of the main text [here Eq.~\eqref{proof3-7}]. The second and the third proofs directly lead to the decomposition of entropy production in Eq.~(6) with the integral fluctuation theorems in Eq.~(7).

Before going into the proofs it is first convenient to recall that the probability of a trajectory $\gammatau = \{n(0); \R_0^\tau; n(\tau)\}$ starting in an eigenstate $\ket{\pi_{n(0)}}$ of the steady state $\pi$ with eigenvalue $\pi_{n(0)}$, is denoted by $P(\gammatau)$. 
According to Born's rule, this probability can be written as $P(\gammatau) = P[n(\tau); \R_0^\tau | n(0)]~ \pi_{n(0)}$, where the conditional probability reads $P[n(\tau); \R_0^\tau | n(0)] = |\bra{\pi_{n(\tau)}} \Li_0^\tau \ket{\pi_{n(0)}}|^2$.
Here we introduced $\Li_0^\tau$ as the operator generating the normalized wavefunction 
\begin{equation}\label{eq:top}
\ket{\psi(\tau)} = \frac{\Li_0^\tau \ket{\pi_{n(0)}}}{\sqrt{\langle \Li_0^{\tau \dagger} \Li_0^\tau \rangle_{\pi_{n(0)}}}}, 
\end{equation}
corresponding to the record $\R_0^\tau$, which verifies the stochastic Sch\"odinger equation~(1) in the main text. From Eq.~\eqref{eq:top} we notice that the relation $P[n(\tau); \R_0^\tau | n(0)] = |\langle \pi_{n(\tau)}| \psi(\tau) \rangle|^2 \langle \Li_0^{\tau \dagger} \Li_0^\tau \rangle_{\pi_{n(0)}}$ is verified.
Analogously, for the probability of the backward trajectory $\tilde P (\tilde{\gamma}_{\{0,\tau\}})$ starting in $\Theta \ket{\pi_{n(\tau)}}$, with $\Theta$ the anti-unitary time-reversal operator, we have $\tilde{P}(\tilde{\gamma}_{\{0,\tau\}}) = \tilde{P}[n(0) \tilde{\R}_\tau^0 | n(\tau)]~ \pi_{n(\tau)}$. 
Importantly, the conditional probabilities for forward and time-reversed trajectories obey the following detailed-balance relation~\cite{horowitzs,PRX}:
\begin{equation} \label{eq:db}
 P[n(\tau); \R_0^\tau | n(0)] = \tilde{P}[n(0) \tilde{\R}_\tau^0 | n(\tau)] ~e^{\sum_{k_j} \Delta S_\mathrm{env}^{k_j}(\R_0^\tau)},
\end{equation}
where $\sum_{k_j} \Delta S_\mathrm{env}^{k_j}(\R_0^\tau)$ is the total entropy change in the environment along the trajectory $\gammatau$, and thus associated with the environmental measurement record $\R_0^\tau$, appearing in Eq.~(2) of the main text.

As already mentioned in the main text, we state the martingale property using conditional averages $\langle X(t) | \gamma_{[0,\tau]} \rangle = \sum_{n(t)} \sum_{\R_\tau^t} X(t) P(\gamma_t | \gamma_{[0,\tau]})$ with respect to ensembles of trajectories $\gamma_{[0, \tau]} \equiv \bigcup_{s=0}^\tau \gamma_{\{0,s\}}$, including all outcomes $n(s)$ of trajectories eventually stopped at all intermediate times in the interval $[0, \tau]$. 
Importantly, the conditional probability in the previous expression is defined as $P(\gamma_{\{0,t\}} | \gamma_{[0,\tau]}) \equiv P(\gamma_{\{0,t\}}, \gamma_{[0,\tau]})/P(\gamma_{[0,\tau]})$ and verifies the following set of equalities: 
\begin{align}\label{cond1}
 P(\gamma_{\{0,t\}} | \gamma_{[0,\tau]}) &\equiv \frac{P(\gamma_{\{0,t\}}, \gamma_{[0,\tau]})}{P(\gamma_{[0,\tau]})} = \frac{P(\gamma_{\{0,t\}}) P(\gamma_{[0,\tau]}|\gamma_{\{0,t\}})}{P(\gamma_{[0,\tau]})} \\ \label{cond2}
 &= \frac{P(\gamma_{\{0,t\}})P(\gamma_{\{0,\tau\}}|\gamma_{\{0,t\}})}{P(\gamma_{\{0,\tau\}})} \\ \label{cond3}
 &= P(\gamma_{\{0,t\}} | \gamma_{\{0,\tau\}}).
 \end{align} 
Here in \eqref{cond1} we used Bayes' theorem to swap conditions, that is $P(\gamma_{\{0,t\}}, \gamma_{[0,\tau]}) = P(\gamma_{\{0,t\}})P(\gamma_{[0,\tau]}|\gamma_{\{0,t\}})$. In \eqref{cond2} we used that the probabilities of virtual measurements at intermediate times in $\gamma_{[0,\tau]}$ are independent, i.e. $P(\gamma_{[0,\tau]}) = P(\gamma_{\{0,\tau\}}) \Pi_{s=0}^\tau |\langle \pi_{n(s)}| \psi(s) \rangle|^2$ 
and $P(\gamma_{[0,\tau]}|\gamma_{\{0,t\}}) = P(\gamma_{\{0,\tau\}}|\gamma_{\{0,t\}}) \Pi_{s=0}^\tau|\langle \pi_{n(s)}| \psi(s) \rangle|^2$, where in both cases $|\langle \pi_{n(s)}| \psi(s) \rangle|^2$ is the probability that the stochastic wavefunction following a trajectory $\gamma_{\{0,t\}}$ at time $s$, $\ket{\psi(s)}$, is found to be in state $\ket{\pi_{n(s)}}$. 
Finally, we used again Bayes' rule swap back conditions and obtain the final expression in \eqref{cond3}.

The above set of equalities \eqref{cond1}-\eqref{cond3} guarantees that arbitrary conditional averages of stochastic functionals $X(t)$ along trajectories $\gammat$ with respect to the ensemble of trajectories $\gamma_{[0, \tau]}$, are exactly equivalent to conditional averages with respect to single trajectories $\gammat$ (without the intermediate virtual measurements): 
\begin{equation}
 \langle X(t) | \gamma_{[0,\tau]} \rangle \equiv \sum_{n(t)} \sum_{\R_\tau^t} X(t) P(\gammat | \gamma_{[0,\tau]}) = \sum_{n(t)} \sum_{\R_\tau^t} X(t) P(\gammat | \gamma_{\{0,\tau\}}) \equiv \langle X(t) | \gamma_{\{0,\tau\}} \rangle,
\end{equation}
as stated in the main text.

\

{\bf Proof 1.}~We now prove the non-martingality of $e^{-\Delta S_\mathrm{tot}(t)}$:
\begin{align} \label{proof1-1}
 \langle e^{-\Delta S_\mathrm{tot}(t)} | \gamma_{[0,\tau]} \rangle &\equiv  \sum_{n(t)} \sum_{\R_\tau^t} e^{-\Delta S_\mathrm{tot}(t)} P(\gammat | \gamma_{[0,\tau]}) =  \sum_{n(t)} \sum_{\R_\tau^t} e^{-\Delta S_\mathrm{tot}(t)} \frac{P(\gamma_{\{0,t\}})}{P(\gamma_{\{0,\tau\}})} |\langle \pi_{n(\tau)}| \psi(\tau) \rangle|^2 \\  \label{proof1-2}
 &= \sum_{n(t)} \sum_{\R_\tau^t} \frac{\tilde{P}(\tilde{\gamma}_{\{0,t\}})}{P(\gamma_{\{0,\tau\}})}  |\langle \pi_{n(\tau)}| \psi(\tau) \rangle|^2 = \sum_{n(t)} \sum_{\R_\tau^t} e^{-\Delta S_\mathrm{tot}(\tau)} \frac{\tilde{P}(\tilde{\gamma}_{\{0,t\}})}{\tilde{P}(\tilde{\gamma}_{\{0,\tau\}})}  |\langle \pi_{n(\tau)}| \psi(\tau) \rangle|^2 \\ \label{proof1-3}
 & = e^{-\Delta S_\mathrm{tot}(\tau)} \sum_{k(\tau)}  \frac{\tilde{P}[n(0); \tilde\R_\tau^0 | k(\tau)] ~\pi_{k(\tau)}}{\tilde{P}(\tilde{\gamma}_{\{0,\tau\}})}  |\langle \pi_{n(\tau)}| \psi(\tau) \rangle|^2 \\  \label{proof1-4}
 &=  e^{-\Delta S_\mathrm{tot}(\tau)} \sum_{k(\tau)}   \frac{P[k(\tau);\R_0^\tau | n(0)] ~\pi_{k(\tau)}}{P[n(\tau);\R_0^\tau | n(0)] ~\pi_{n(\tau)}}  |\langle \pi_{n(\tau)}| \psi(\tau) \rangle|^2 \\ \label{proof1-5}
 &= e^{-\Delta S_\mathrm{tot}(\tau)} \sum_{k(\tau)}   \frac{|\langle \pi_{k(\tau)}| \psi(\tau) \rangle|^2 ~\pi_{k(\tau)}}{\pi_{n(\tau)}} \\ \label{proof1-6}
 &= e^{-\Delta S_\mathrm{tot}(\tau) + \Delta S_\mathrm{unc}(\tau)}. \qquad  \qquad  \qquad  \qquad  \qquad  \qquad  \qquad  \qquad \qquad \qquad \qquad \qquad \square
\end{align}
In the second equality of~\eqref{proof1-1} we used Eq.~\eqref{cond3} and that $P(\gamma_{\{0,\tau\}}|\gamma_{\{0,t\}}) = |\langle \pi_{n(\tau)}| \psi(\tau) \rangle|^2$ is the probability that the stochastic wavefunction, $\ket{\psi(\tau)}$, following the trajectory $\gamma_{\{0,t\}}$, is found to be in state $\ket{\pi_{n(\tau)}}$ at time $\tau$. 
In the first part of Eq.~\eqref{proof1-2} we introduced the expression of the total entropy production $\Delta S_\mathrm{tot}(t)$ as given in Eq.~(2) of the main text, leading to $e^{-\Delta S_\mathrm{tot}(t)} = \tilde{P}(\tilde{\gamma}_{\{0,t\}}) / P(\gamma_{\{0,t\}})$, while in the second part the same expression for $\Delta S_\mathrm{tot}(\tau)$ is used.
Then, in~\eqref{proof1-3} we used the marginalization property $\sum_{n(t)} \sum_{\R_\tau^t} \tilde{P}(\tilde{\gamma}_{\{0,t\}}) = \sum_{k(\tau)} \tilde{P}(\tilde{\gamma}_{\{0,\tau\}}) = \sum_{k(\tau)} \tilde{P}[n(0); \tilde\R_\tau^0 | k(\tau)] ~\pi_{k(\tau)}$, following from the stationarity of the dynamics in the time-reversed processes. 
In Eq.~\eqref{proof1-4}, we split the probability $\tilde{P}(\tilde{\gamma}_{\{0,\tau\}})$ in the denominator and used the detailed-balance relation~\eqref{eq:db} in both numerator and denominator. In \eqref{proof1-5} we use that $P[k(\tau); \R_0^\tau | n(0)] = |\langle \pi_{k(\tau)}| \psi(\tau) \rangle|^2 \langle \Li_0^{\tau \dagger} \Li_0^\tau \rangle_{\pi_{n(0)}}$ 
and $P[n(\tau); \R_0^\tau | n(0)] = |\langle \pi_{n(\tau)}| \psi(\tau) \rangle|^2 \langle \Li_0^{\tau \dagger} \Li_0^\tau \rangle_{\pi_{n(0)}}$.
Finally, in Eq.~\eqref{proof1-6} we recognize the uncertainty entropy production $\Delta S_\mathrm{unc} (t)=\ln[\sum_{k(\tau)} |\langle \pi_{k(\tau)}| \psi(\tau) \rangle|^2 ~\pi_{k(\tau)}/\pi_{n(\tau)}] = - \ln[\pi_{n(\tau)}/ \langle \pi \rangle_{\psi(t)}]$ as given in Eq.~(3) of the main text. 
This demonstrates that the exponential of $\Delta S_\mathrm{tot}(t)$ is not a martingale in general, but only in the classical limit  when $\Delta S_\mathrm{unc}(\tau) = 0$.
In that limit we therefore recover the results derived in Ref.~\cite{Neris}.

\

{\bf Proof 2.}~Now we proof the generalized integral fluctuation theorem for the uncertainty entropy production, $\langle e^{-\Delta S_\mathrm{unc}(t)} | \gamma_{[0,\tau]} \rangle = 1$, used for deriving the decomposition of entropy production in the main text [Eqs.~(6)-(8)]:
\begin{align} \label{proof2-1}
  \langle e^{-\Delta S_\mathrm{unc}(t)} | \gamma_{[0,\tau]} \rangle &\equiv \sum_{n(t)} \sum_{\R_\tau^t} e^{-\Delta S_\mathrm{unc}(t)} P(\gammat | \gamma_{[0,\tau]}) = \sum_{n(t)} \sum_{\R_\tau^t} e^{-\Delta S_\mathrm{unc}(t)} \frac{P(\gamma_{\{0,t\}})}{P(\gamma_{\{0,\tau\}})} |\langle \pi_{n(\tau)}| \psi(\tau) \rangle|^2 \\ \label{proof2-2}
  &= \sum_{n(t)} \sum_{\R_\tau^t}   \frac{|\langle \pi_{n(\tau)}| \psi(\tau) \rangle|^2 \pi_{n(t)}}{\sum_{l(t)}  |\langle \pi_{l(t)}| \psi(t) \rangle|^2 \pi_{l(t)}} \frac{P(\gamma_{\{0,t\}})}{P(\gamma_{\{0,\tau\}})}  \\  \label{proof2-3}
  &=\sum_{n(t)} \sum_{\R_\tau^t}   \frac{|\langle \pi_{n(t)}| \psi(t) \rangle|^2 \pi_{n(t)}}{\sum_{l(t)}  |\langle \pi_{l(t)}| \psi(t) \rangle|^2 \pi_{l(t)}} \frac{\langle \Li_0^{t \dagger} \Li_0^{t} \rangle_{\pi_{n(0)}}}{\langle \Li_0^{\tau \dagger} \Li_0^{\tau} \rangle_{\pi_{n(0)}}} =  \sum_{\R_\tau^t} \frac{\langle \Li_0^{t \dagger} \Li_0^{t} \rangle_{\pi_{n(0)}}}{\langle \Li_0^{\tau \dagger} \Li_0^{\tau} \rangle_{\pi_{n(0)}}} = 1. \qquad \qquad \square
\end{align}
Again in \eqref{proof2-1} we used Eq.~\eqref{cond2} and $P(\gamma_{\{0,\tau\}}|\gamma_{\{0,t\}}) = |\langle \pi_{n(\tau)}| \psi(\tau) \rangle|^2$ to reach the second equality, and \eqref{proof2-2} follows by substituting the expression of $\Delta S_\mathrm{unc}(t)$ [Eq.~(3) in the main text]. 
In the first part of \eqref{proof2-3} we split the probabilities $P(\gammat)= p[n(t);\R_0^t | n(0)] \pi_{n(0)}$ and $P(\gammatau) =  p[n(\tau);\R_0^\tau | n(0)] \pi_{n(0)}$, and used the relation $P[n(t); \R_0^\tau | n(0)] = |\langle \pi_{n(t)}| \psi(t) \rangle|^2 \langle \Li_0^{t \dagger} \Li_0^t \rangle_{\pi_{n(0)}}$. Summing over the index $n(t)$ yields to the second equality in \eqref{proof2-3}, 
which yields $1$ upon making the last sum over the environmental measurement record $\R_\tau^t$ in the interval $[\tau,t]$. 
Notice that when taking $\tau = 0$, the above equation reduces to the integral fluctuation theorem $\langle e^{-\Delta S_\mathrm{unc} (t)} \rangle = 1$ [Eq.~(7) in the main text].

\

{\bf Proof 3.}~Finally we provide the proof for the martingality of $e^{-\Delta S_\mathrm{mar}(t)}$, Eq.~(5) in the main text:
\begin{align} \label{proof3-1}
 \langle e^{-\Delta S_\mathrm{mar}(t)} | \gamma_{[0,\tau]} \rangle &\equiv  \sum_{n(t)} \sum_{\R_\tau^t} e^{-\Delta S_\mathrm{mar}(t)} P(\gammat | \gamma_{[0,\tau]}) = \sum_{n(t)} \sum_{\R_\tau^t} e^{-\Delta S_\mathrm{tot}(t) + \Delta S_\mathrm{unc}(t)} \frac{P(\gamma_{\{0,t\}})}{P(\gamma_{\{0,\tau\}})} |\langle \pi_{n(\tau)}| \psi(\tau) \rangle|^2 \\   \label{proof3-2}
 &= \sum_{n(t)} \sum_{\R_\tau^t} \sum_{l(t)} \frac{|\langle \pi_{l(t)}| \psi(t) \rangle|^2 \pi_{l(t)}}{\pi_{n(t)}} \frac{\tilde{P}(\tilde{\gamma}_{\{0,t\}})}{P(\gamma_{\{0,\tau\}})}  |\langle \pi_{n(\tau)}| \psi(\tau) \rangle|^2 \\ \label{proof3-3}
 &= e^{-\Delta S_\mathrm{tot}(\tau)} \sum_{n(t)} \sum_{\R_\tau^t} \sum_{l(t)} \frac{|\langle \pi_{l(t)}| \psi(t) \rangle|^2 \pi_{l(t)}}{\pi_{n(t)}} \frac{\tilde{P}(\tilde{\gamma}_{\{0,t\}})}{\tilde{P}(\tilde{\gamma}_{\{0,\tau\}})}  |\langle \pi_{n(\tau)}| \psi(\tau) \rangle|^2 \\ \label{proof3-4}
 &= e^{-\Delta S_\mathrm{tot}(\tau)} \sum_{n(t)} \sum_{\R_\tau^t} \sum_{l(t)} \frac{P[l(t); \R_0^t|n(0)] \pi_{l(t)}}{\langle \Li_0^{t \dagger} \Li_0^t  \rangle_{\pi_{n(0)}}}  \frac{\tilde{P}[n(0); \tilde \R_t^0|n(t)]}{\tilde{P}(\tilde{\gamma}_{\{0,\tau\}})}  |\langle \pi_{n(\tau)}| \psi(\tau) \rangle|^2  \\ \label{proof3-5}
 &= e^{-\Delta S_\mathrm{tot}(\tau)} \sum_{l(t)} \sum_{\R_\tau^t} \frac{\tilde P[n(0); \tilde \R_t^0|l(t)] \pi_{l(t)}}{\tilde{P}(\tilde{\gamma}_{\{0,\tau\}})}  |\langle \pi_{n(\tau)}| \psi(\tau) \rangle|^2  \\ \label{proof3-6}
 &= e^{-\Delta S_\mathrm{tot}(\tau)} \sum_{k(\tau)}  \frac{\tilde{P}[n(0); \tilde\R_\tau^0 | k(\tau)] ~\pi_{k(\tau)}}{\tilde{P}(\tilde{\gamma}_{\{0,\tau\}})}  |\langle \pi_{n(\tau)}| \psi(\tau) \rangle|^2 \\ \label{proof3-7}
 &= e^{-\Delta S_\mathrm{mar}(\tau)}. \qquad  \qquad  \qquad  \qquad  \qquad  \qquad  \qquad  \qquad \qquad \qquad \qquad \qquad \qquad \qquad \qquad \square
\end{align}
As before, we used in the first line \eqref{proof3-1} the equality \eqref{cond2} and $P(\gamma_{\{0,\tau\}}|\gamma_{\{0,t\}}) = |\langle \pi_{n(\tau)}| \psi(\tau) \rangle|^2$, together with the definition of the martingale entropy production $\Delta S_\mathrm{mar}(t) = \Delta S_\mathrm{tot}(t) - \Delta S_\mathrm{unc}(t)$.
Then using $e^{-\Delta S_\mathrm{tot}(t)} = \tilde{P}(\tilde{\gamma}_{\{0,t\}}) / P(\gamma_{\{0,t\}})$, and the definition of $\Delta S_\mathrm{unc}(t)$ [Eq.~(3) in main text], we reach \eqref{proof3-2}. In line \eqref{proof3-3} we introduced $e^{-\Delta S_\mathrm{tot}(\tau)} = \tilde{P}(\tilde{\gamma}_{\{0,\tau\}}) / P(\gamma_{\{0,\tau\}})$. 
In \eqref{proof3-4} we used $|\langle \pi_{l(t)}| \psi(t) \rangle|^2  = P[l(t); \R_0^t|n(0)]/\langle \Li_0^{t \dagger} \Li_0^t  \rangle_{\pi_{n(0)}}$ and split the probability $\tilde{P}(\tilde{\gamma}_{\{0,t\}}) = \tilde{P}[n(0); \tilde \R_t^0|n(t)] \pi_{n(t)}$. Crucially, in order to obtain \eqref{proof3-5} we applied the detailed-balance relation in Eq.~\eqref{eq:db} to both 
$P[l(t); \R_0^t|n(0)]$ and $\tilde{P}[n(0); \tilde \R_t^0|n(t)]$, and performed the sum over $n(t)$ leading to $\sum_{n(t)} {P}[n(t); \R_0^t|n(0)] = \langle \Li_0^{t \dagger} \Li_0^t  \rangle_{\pi_{n(0)}}$. Summing \eqref{proof3-5} over $l(t)$ and the measurement record $\R_\tau^t$ leads again to 
the marginalization $\sum_{l(t)} \sum_{\R_\tau^t}  \tilde{P}[n(0); \tilde\R_t^0 | l(t)] ~\pi_{l(t)} = \sum_{l(t)} \sum_{\R_\tau^t} \tilde{P}(\tilde{\gamma}_{\{0,t\}}) = \sum_{k(\tau)} \tilde{P}(\tilde{\gamma}_{\{0,\tau\}}) = \sum_{k(\tau)} \tilde{P}[n(0); \tilde\R_\tau^0 | k(\tau)] ~\pi_{k(\tau)}$, which results on Eq.~\eqref{proof3-6}. Finally, noticing that \eqref{proof3-6} is identically equal to Eq.~\eqref{proof1-3}, 
we obtain the final result in \eqref{proof3-7}, concluding the proof.

Since Eqs.~\eqref{proof3-1}-\eqref{proof3-7} are verified and $e^{-\Delta S_\mathrm{mar}(t)} < \infty$ is bounded, we conclude that $\Delta S_\mathrm{mar}(t)$ is an exponential martingale. 
Again, choosing $\tau = 0$ we recover from Eq.~\eqref{proof3-7} the integral fluctuation theorem $\langle  e^{-\Delta S_\mathrm{mar}(t)} \rangle = 1$ [Eq.~(7) in the main text].
Applying Jensen's inequality, $e^{\langle x \rangle} \leq \langle e^{x} \rangle$, to the two integral fluctuation theorems in Eq.~(7) for the uncertainty and martingale entropy productions, we obtain the second-law-like inequalities $\langle \Delta S_\mathrm{unc}(t) \rangle \geq 0$, and $\langle \Delta S_\mathrm{mar}(t) \rangle \geq 0$ from which the inequality~(8) of the main text follows:
\begin{equation}
 \langle \Delta S_\mathrm{tot}(t) \rangle = \langle \Delta S_\mathrm{unc}(t) \rangle + \langle \Delta S_\mathrm{mar}(t) \rangle \geq \langle \Delta S_\mathrm{mar}(t) \rangle.
\end{equation}
We also notice that the bound $\langle \Delta S_\mathrm{tot}(t) \rangle \geq  \langle \Delta S_\mathrm{unc}(t) \rangle$ follows in the same way.

\section{Proof of the stopping-time fluctuation theorem} \label{s2}

In this section we provide a proof of the stopping-time fluctuation theorem in Eq.~(9) of the main text. The proof is based on Doob's optional stopping theorem~\cite{Doobs}, which holds for processes $M(t)$ defined in a probability space $(\Omega, \mathcal{F}, \mathds{P})$, that are Martingales. Here $\Omega$ is the sample space, $\mathcal{F}$ is a $\sigma-$algebra, and $\mathds{P}$ a probability measure. 
In our context,  $\mathcal{F}$ is the set of all possible trajectories as specified by $\gamma_{\{0, t \}}$, representing different events in the sample space of events $\Omega$. Moreover, the probability measure $\mathds{P}$ associates a probability $P(\gammat)$ to any element $\gammat$ of the $\sigma-$algebra $\mathcal{F}$.
We also introduce $\mathcal{T}$ as a bounded stopping time, i.e. $\mathcal{T} < c$ for some arbitrary constant $c$, or for uniformly integrable process, $|M(s)| < c^\prime$, for $s \equiv \mathrm{min}(\mathcal{T}, t)$ and some arbitrary constant $c^\prime$~\cite{Williams}.

Here we use the martingale $M(t)=e^{-\Delta S_\mathrm{mar}(t)}$, and we define a stopping time $\mathcal{T}$ by introducing the \emph{filtrations} (or subsets) $F_\tau \subseteq \mathcal{F}$, here defined as sets of all possible trajectories $\gammat$ for which a given ``stopping'' condition is satisfied for the first time at time $\tau$. Therefore, $M(\tau) = M(\mathcal{T})$ for all trajectories in $F_\tau$. 
Doob's optional stopping theorem can be proved for stopping times obeying $\mathcal{T} \leq t$:
\begin{align}\label{stopping}
 \langle M(\mathcal{T}) \rangle = \sum_{\tau=0}^{t} \sum_{\gamma_{\{0,\tau\}} \in F_\tau} P(\gamma_{\{0, \tau \}}) M(\tau) = \sum_{\tau=0}^{t} \sum_{\gamma_{\{0,\tau\}} \in F_\tau}  P(\gamma_{\{0, \tau \}}) \langle M(t) | \gamma_{[0,\tau]} \rangle = \langle M(t) \rangle = \langle M(0) \rangle. \qquad \square
\end{align}
In Eq.~\eqref{stopping}, the crucial step  from the first to the second equality is the use of the Martingale property $ M(\tau) = \langle M(t) | \gamma_{[0,\tau]} \rangle$. The subsequent equality follows from the fact that we are summing over all  sets $F_\tau$ of trajectories verifying the condition at different times, that is, all trajectories (we recall that $\mathcal{T} \leq t$). 
The last equality follows from the fact that martingale processes have no drift and thus $\langle M(t) \rangle = \langle M(0) \rangle$.
Finally, by noticing that $\langle M(t) \rangle= \langle e^{- \Delta S_\mathrm{mar}(t)} \rangle = \langle M(0) \rangle = 1$ we obtain from \eqref{stopping}:
\begin{equation}
 \langle e^{- \Delta S_\mathrm{mar}(\mathcal{T})} \rangle = 1,
\end{equation}
which proves Eq.~(9) in the main text. We remark that in the above proof we use sums over discrete times $\tau$, but in the continuous-time limit these averages have to be understood as  path integrals~\cite{Shamik}. 
Note that in Eq.~\eqref{stopping} the stopping times $\mathcal{T}$ need to occur in the interval $[0,t]$ for any arbitrary finite $t$. However, following \cite{Williams}, we may also take $t \rightarrow \infty$ whenever $|M(\tau)|$ is finite for all $\tau \equiv \mathrm{min}(\mathcal{T}, t)$.
The stopping-time second law inequality in Eq.~(10) follows by applying Jensen's inequality to the above equation, that is, $e^{\langle -\Delta S_\mathrm{mar}(\mathcal{T}) \rangle} \leq \langle e^{-\Delta S_\mathrm{mar}(\mathcal{T})} \rangle = 1$, which implies $\langle \Delta S_\mathrm{mar}(\mathcal{T}) \rangle \geq 0$.

\section{Proof of Doob's maximal inequality and extreme-value bounds} \label{s3}

In this section we prove Eqs.~(11) and~(12) in the main text concerning  finite-time infima statistics. We make use of Doob's maximal inequality~\cite{Doobs}, bounding the probability of the supremum of the positive martingale process $M(t) = e^{-\Delta S_\mathrm{mar}(t)}$. 
We define sets $F^\prime_\tau \in \mathcal{F}$ now characterizing the trajectories for which  a value  $\lambda \geq 0$ is first reached at time $s$, that is, $M(s) <  \lambda$ for $s\leq \tau$ and  $M(\tau) \geq  \lambda$. Doob's maximal inequality follows as (see also Ref.~\cite{Williams}):
\begin{align} \label{maximal}
 \mathsf{Pr}(\mathrm{sup}_{\tau \in [0,t]} M(\tau) \geq \lambda) &= \sum_{\tau=0}^t \sum_{\gamma_{\{0,\tau\}} \in F^\prime_\tau} P(\gamma_{\{0, \tau \}}) ~ \leq ~ \sum_{\tau=0}^t \sum_{\gamma_{\{0,\tau\}}  \in F^\prime_\tau} \frac{M(\tau)}{\lambda} P(\gamma_{\{0, \tau \}}) \\ \label{maximal2}
 &= \sum_{\tau=0}^t \sum_{\gamma_{\{0, \tau\}}  \in F^\prime_\tau} \frac{\langle M(t) | \gamma_{[0, \tau]} \rangle}{\lambda} = \frac{\langle M(t) \rangle}{\lambda},  \qquad  \qquad \qquad \qquad  \qquad\qquad \square
\end{align}
and therefore $\mathsf{Pr}(\mathrm{sup}_{\tau \in [0,t]} M(\tau) \geq \lambda) \leq \langle M(t) \rangle / \lambda$. The inequality in \eqref{maximal} follows from the fact that $M(\tau) \leq \lambda$ by construction of the sets $F^\prime_\tau$, and again the crucial step from \eqref{maximal} to \eqref{maximal2} has been to introduce the Martingale property. 
To prove Eq.~(11) then we note that $\mathrm{sup}_{\tau \in [0,t]} M(\tau) \geq \lambda$ is equivalent to 
$\mathrm{inf}_{\tau \in [0,t]} \Delta S_\mathrm{mar}(\tau) \leq - \ln \lambda \equiv - \xi$. Therefore $\mathsf{Pr}(\mathrm{sup}_{\tau \in [0,t]} M(\tau) \geq \lambda) = \mathsf{Pr}(\mathrm{inf}_{\tau \in [0,t]} \Delta S_\mathrm{mar}(\tau)  \leq - \xi)$, and we can rewrite Eq.~\eqref{maximal2} as:
\begin{equation}
 \mathsf{Pr}(\mathrm{inf}_{\tau \in [0,t]} \Delta S_\mathrm{mar}(\tau) \leq - \xi) \leq \frac{\langle e^{- \Delta S_\mathrm{mar}(t)}\rangle}{e^{ \xi}} = e^{-\xi},
\end{equation}
where in the last line we used the integral fluctuation theorem $\langle e^{- \Delta S_\mathrm{mar}(t)}\rangle = 1$.

Finally, following Ref.~\cite{Neris}, we have that Eq.~(11) directly implies a bound on the average finite-time infimum of $\Delta S_\mathrm{mar}(t)$, reading $\left\langle\inf_{\tau\in[0,t]} \Delta S_\mathrm{mar}(\tau)  \right\rangle\geq  -1$.
Now using that $\Delta S_\mathrm{mar}(t) = \Delta S_\mathrm{tot}(t) - \Delta S_\mathrm{unc}(t)$ and that the uncertainty entropy production is bounded by $\ln[\pi_\mathrm{min}/\pi_\mathrm{max}] \leq \Delta S_\mathrm{unc} \geq \ln[\pi_\mathrm{max}/\pi_\mathrm{min}]$, we obtain Eq.~(12) in the main text:
\begin{align}
 \left\langle\inf_{\tau\in[0,t]} \Delta S_\mathrm{tot}(\tau)   \right\rangle +  \ln \left[\frac{\pi_\mathrm{max}}{\pi_\mathrm{min}}\right] \geq \left\langle\inf_{\tau\in[0,t]} \Delta S_\mathrm{mar}(\tau)  \right\rangle \geq  -1.
\end{align}


\begin{thebibliography}{99}


\bibitem{seifert}
U. Seifert, {\it Entropy production along a Stochastic Trajectory and a Integral Fluctuation Theorem}, Phys. Rev. Lett. {\bf 95}, 040602 (2005);

\bibitem{lebowitz}
J. L. Lebowitz, and H. Spohn,  {\it A Gallavotti-Cohen-type symmetry in the large deviation functional for stochastic dynamics}, J. Stat. Phys. {\bf 95}, 333-365  (1999).

\bibitem{sekimoto} K. Sekimoto, {\it Stochastic energetics} (Springer, Berlin, 2010).

\bibitem{SeifertREV}
U. Seifert, {\it Stochastic thermodynamics, fluctuation theorems and molecular machines}, Rep. Prog. Phys., {\bf 75}, 126001 (2012).

\bibitem{JarzynskiREV}
C. Jarzynski, {\it Equalities and Inequalities: Irreversibility and the Second Law of Thermodynamics at the Nanoscale}, Annu. Rev. Condens. Matter Phys. {\bf 2}, 329-351 (2011).

\bibitem{quantum} 
J. Goold, M. Huber, A. Riera, L. del Rio, P. Skrzypczyk, {\it The role of quantum information in thermodynamics -- a topical review}, J. Phys. A: Math. Theor. {\bf 49}, 143001 (2016);
S. Vinjanampathy, and J. Anders, {\it Quantum Thermodynamics}, Contemp. Phys. {\bf 57}, 545-579 (2016).

\bibitem{fluctuations1} M. Esposito, U. Harbola, and S. Mukamel, {\it Nonequilibrium fluctuations, fluctuation theorems, and counting statistics in quantum systems}, Rev. Mod. Phys., {\bf 81}, 1665 (2009);

\bibitem{fluctuations2} M. Campisi, P. H\"anggi, and P. Talkner, {\it Colloquium: Quantum fluctuation relations: Foundations and applications}, Rev. Mod. Phys., {\bf 83}, 771 (2011).

\bibitem{deffner} S. Deffner and E. Lutz, {\it Nonequilibrium Entropy Production for Open Quantum Systems}, Phys. Rev. Lett., {\bf 107}, 140404 (2011).

\bibitem{tasaki} Y. Morikuni, and H. Tasaki, {\it Quantum Jarzynski-Sagawa-Ueda Relations}, J. Stat. Phys. {\bf 143}, 1-10 (2011).

\bibitem{ueda} K. Funo, Y. Murashita, and M. Ueda, {\it Quantum nonequilibrium equalities with absolute irreversibility}, New J. Phys. {\bf 17}, 075005 (2015).

\bibitem{manzano} G. Manzano, J. M. Horowitz, and J. M. R. Parrondo, {\it Quantum Fluctuation Theorems for Abitrary Environments: Adiabatic and Non-adiabatic Entropy Production}, Phys. Rev. X {\bf 8}, 031037 (2018).

\bibitem{bartolotta} A. Bartolotta and S. Deffner, {\it Jarzynski Equality for Driven Quantum Field Theories}, Phys. Rev. X {\bf 8}, 011033 (2018).


\bibitem{experimentalFT1}
T. B. Batalh\~ao, A. M. Souza, L. Mazzola, R. Auccaise, R. S. Sarthour, I. S. Oliveira, J. Goold, G. De Chiara, M. Paternostro, and R. M. Serra, {\it Experimental Reconstruction of Work Distribution 
and Study of Fluctuation Relations in a Closed Quantum System}, Phys. Rev. Lett. {\bf 113}, 140601 (2014);

\bibitem{experimentalFT2}
S. An, J.-N. Zhang, M. Um, D. Lv, Y. Lu, J. Zhang, Z.-Q. Yin, H. T. Quan, and K. Kim, {\it Experimental test of the quantum Jarzynski equality with a trapped-ion system}, Nat. Phys. {\bf 11}, 193 (2015).


\bibitem{demons1} J. M. R. Parrondo, J. M. Horowitz, and T. Sagawa, {\it Thermodynamics of information}, Nature Phys. {\bf 11}, 131 (2015).

\bibitem{demons2} H. Leff, A. F. Rex (Eds.), {\it Maxwell's Demon 2: Entropy, Classical and Quantum Information, Computing}, (CRC Press, 2002).

\bibitem{martingale0} A martingale process is a stochastic process defined on a probability space whose expected value at any time $t$ equals its value at some previous time $\tau<t$ when conditioned on observations up to that time $\tau$. 
More formally, $M(t)$ is a martingale if it is bounded $\langle M(t) \rangle < \infty$ for all $t$, and verifies $\langle M(t) | M_{\{0,\tau\}} \rangle = M(\tau)$, where the later average is conditioned on all the previous values  $M_{\{0,\tau\}}$ of the process up to time $\tau$~\cite{martingales}.


\bibitem{martingales} D. Williams, {\it Probability with Martingales}, (Cambridge University Press, Cambridge, 1991).

\bibitem{economy} S. Pliska,  {\it Introduction to mathematical finance}. (Blackwell publishers, Oxford, 1997).

\bibitem{neri} I. Neri, \'E. Rold\'an, and F. J\"ulicher, {\it Statistics of Infima and Stopping Times of Entropy Production and Applications to Active Molecular Processes}, Phys. Rev. X {\bf 7}, 011019 (2017). 

\bibitem{raphaelshamik} R. Chetrite and S. Gupta, {\it Two refreshing views of fluctuation theorems through kinematics elements and exponential martingale}, J. Stat. Phys., {\bf 143}, 543 (2011).

\bibitem{haoge} H. Ge, X. Jin, {\it Anomalous contribution and fluctuation theorems in singular perturbed diffusion processes}, arXiv:1811.04529 (2018).

\bibitem{ventejou} B. Vent\'ejou, K. Sekimoto, {\it Progressive quenching: Globally coupled model}, Phys. Rev. E, {\bf 97}, 062150 (2018).

\bibitem{singh} S. Singh, {\it et al.}, {\it Extreme reduction of entropy  in an electronic double dot} arXiv:1712.01693 (2017).

\bibitem{chetrite} R. Chetrite, S. Gupta, I. Neri, and \'E. Rold\'an, {\it Martingale theory for housekeeping heat}, EPL {\bf 124}, 60006 (2018).

\bibitem{pigolotti} S. Pigolotti, I.  Neri, \'E. Rold\'an, and F. J\"ulicher, {\it Generic properties of stochastic entropy production}, Phys. Rev. Lett. {\bf 119}, 140604 (2017).

\bibitem{bauer} M. Bauer, and D. Bernard, {\it Convergence of repeated quantum nondemolition measurements and wave-function collapse}, Phys. Rev. A {\bf 84}, 044103 (2011).

\bibitem{adler} S. L. Adler, D. C. Brody, T. A. Brun, and L. P. Hughston, {\it Martingale models for quantum state reduction}  J. Phys. A {\bf 34}, 8795 (2001). 


\bibitem{milburn} H. M. Wiseman and G. J. Milburn, {\it Quantum Measurement and Control} (Cambridge University Press, Cambridge, 2010).

\bibitem{trajectories}
M. B. Plenio and P. L. Knight, {\it The quantum-jump approach to dissipative dynamics in quantum optics}, Rev. Mod. Phys. {\bf 70}, 101 (1998).


\bibitem{coherence1} M. O. Scully, M. S. Zubairy, G. S. Agarwal, and H. Walther, {\it Extracting work from a single heat bath via vanishing quantum coherence}, Science {\bf 299}, 862 (2003);

\bibitem{coherence2} M. Lostaglio, D. Jennings, T. Rudolph, {\it Quantum coherence, time-translation symmetry, and thermodynamics}, Phys. Rev. X {\bf 5}, 021001 (2015);

\bibitem{coherence3} P. Krammerlander, and J. Anders, {\it Coherence and measurement in quantum thermodynamics}, Sci. Rep. {\bf 6}, 22174 (2016).

\bibitem{correlations1} L. del Rio, J. \AA berg, R. Renner, O. Dahlsten, and V. Vedral, {\it The thermodynamic meaning of negative entropy}, Nature {\bf 474}, 61-63 (2011).

\bibitem{correlations2} M. N. Bera, A. Riera, M. Lewenstein, A. Winter, {\it Generalized laws of thermodynamics in the presence of correlations}, Nature Commun. {\bf 8}, 2180 (2017);

\bibitem{correlations3} G. Manzano, F. Plastina, and R. Zambrini, {\it Optimal Work Extraction and Thermodynamics of Quantum Measurements and Correlations}, Phys. Rev. Lett. {\bf  121}, 120602 (2018).

\bibitem{horowitz} J. M. Horowitz, {\it Quantum-trajectory approach to the stochastic thermodynamics of a forced harmonic oscillator}, Phys. Rev. E {\bf 85}, 031110 (2012).

\bibitem{horowitz2} J. M. Horowitz, and J.M.R. Parrondo, {\it Entropy production along nonequilibrium quantum jump trajectories}, New. J. Phys. {\bf 15}, 085028 (2013),

\bibitem{manzanoPRE}
G. Manzano, J. M. Horowitz, and J. M. R. Parrondo, {\it Nonequilibrium potential and fluctuation theorems for quantum maps}, Phys. Rev. E {\bf 92}, 032129 (2015).

\bibitem{auffeves} C. Elouard, N. K. Bernardes, A. R. R. Carvalho, M. F. Santos, and A. Auff\'eves, {\it Probing quantum fluctuation theorems in engineered reservoirs}, New J. Phys. {\bf 19}, 103011 (2017).

\bibitem{gheradini} S. Gherardini, M. M. Muller, A. Trombettoni, S.  Ruffo, and F.  Caruso,  {\it Reconstructing quantum entropy production to probe irreversibility and correlations}, Quant. Sci. Technol. {\bf 3}, 035013 (2018).

\bibitem{landi} J. P. Santos, L. C. C\'eleri, G. T. Landi, and M. Paternostro, {\it The role of quantum coherence in non-equilibrium entropy production}, npj Quantum Information {\bf 5}, 23 (2019).  

\bibitem{hekking} F. W. J. Hekking, and J. P. Pekola, {\it Quantum jump approach for work and dissipation in a two-level system}, Phys. Rev. Lett. {\bf 111}, 093602 (2013),

\bibitem{campisi} M. Campisi, J. P. Pekola, and R. Fazio, {\it Nonequilibrium fluctuations in quantum heat engines: theory, example, and possible solid state experiments}, New J. Phys. {\bf 17}, 035012 (2015),

\bibitem{circuits} J. P. Pekola, {\it Towards quantum thermodynamics in electronic circuits}, Nat. Phys. {\bf 11}, 118-123 (2015);

\bibitem{romito} J. J. Alonso, E. Lutz, and A. Romito, {\it Thermodynamics of Weakly Measured Quantum Systems}, Phys. Rev. Lett. {\bf 116}, 080403 (2016).

\bibitem{gong} Z. Gong, Y. Ashida, and M. Ueda, {\it Quantum-trajectory thermodynamics with discrete feedback control}, Phys. Rev. A {\bf 94}, 012107 (2016).

\bibitem{elouard} C. Elouard, D. A. Herrera-Mart\'i, M. Clusel, A. Auff\'eves, {\it The role of quantum measurement in stochastic thermodynamics}, npj Quantum Information {\bf 3}, 9 (2017).




\bibitem{demon} N. Cottet, {\it et al.}, {\it Observing a quantum Maxwell demon at work}, PNAS {\bf 114}, 7561 (2017).

\bibitem{naghiloo} M. Naghiloo, D. Tan, P. M. Harrington, J. J. Alonso, E. Lutz, A. Romito, and K. W. Murch, {\it Thermodynamics along individual trajectories of a quantum bit}, arXiv:1703.05885 (2017).



\bibitem{Lindblad} G. Lindblad,  {\it On the generators of quantum dynamical semigroups},  Comm. Math. Phys. {\bf 48}, 119-130. (1976).

\bibitem{rivas} A. Rivas, and S. F. Huelga, {\it Open Quantum Systems: An Introduction} (Springer, Berlin, 2012).

\bibitem{climit} This happens when $[\pi, L_k] = \alpha_k L_k$ for some set of coefficients $\{\alpha_k \}_{k=1}^K$, and $[\pi, H] = 0$ for $H$ without degenerate gaps. That is, when the Lindblad operators only induce jumps between single eigenstates of $\pi$ (see e.g.~\cite{manzanoPRE}).





\bibitem{supplemental} See Supplemental Material for detailed proofs of our main results.

\bibitem{oono} Y. Oono and M. Paniconi, {\it Steady state thermodynamics},  Prog. Theor. Phys. Suppl., {\bf 130}  29 (1998).

\bibitem{ev} M. Esposito,  C.  Van den Broeck, {\it Three detailed fluctuation theorems}, Phys. Rev. Lett., {\bf 104}  090601 (2010).

\bibitem{harris} R. J. Harris, and G. M. Sch\"utz, {\it  Fluctuation theorems for stochastic dynamics}, J. Stat. Mech. {\bf 2007}(07), P07020 (2007).

\bibitem{VN}  Its average is bounded by $ 0 \leq \langle \Delta S_{\rm unc} (t) \rangle \leq S_{\rm VN}(\pi)$, with $S_{\rm VN}(\pi)\equiv - \tr[\pi \ln \pi]$ the Von Neumman entropy of the steady state. 


\bibitem{doob} J. L. Doob, {\it Stochastic Processes} (John Wiley and Sons, Chapman and Hall, New York, 1953); 
J. L. Doob, {\it Measure Theory} (Springer Science and Business Media, Berlin, 1994), Vol. 143.




\bibitem{qutip} J. R. Johansson, P. D. Nation, and F. Nori, {\it QuTiP: An open-source Python framework for the dynamics of open quantum systems.}, Comp. Phys. Comm. {\bf 183}, 1760-1772 (2012).





\bibitem{Masuyama} Y. Masuyama, {\it et al.} {\it Information-to-work conversion by Maxwell's demon in a superconducting circuit quantum electrodynamical system} Nat. Commun. {\bf 9}, 1291 (2018).  

\bibitem{carmichael} Z. K. Minev, {\it et al.} {\it To catch and reverse a quantum jump mid-flight}, arXiv:1803.00545 (2018). 

\bibitem{Dima} J. V. Koski, A. Kutvonen, I. M. Khaymovich, T.  Ala-Nissila, J. P. Pekola, {\it On-chip Maxwell's demon as an information-powered refrigerator}, Phys. Rev. Lett. {\bf 115}, 260602 (2015).

\bibitem{dicarlo} G. de Lange, {\it et al.} {\it Reversing Quantum Trajectories with Analog Feedback}, Phys. Rev. Lett. {\bf 112}, 080501 (2014).

\bibitem{haroche} C. Sayrin, {\it et al.} {\it Real-time quantum feedback prepares and stabilizes photon number states}, Nature {\bf 477}, 73-77 (2011).

\bibitem{barreiro} J. T. Barreiro, {\it et al.} {\it An open-system quantum simulator with trapped ions}, Nature {\bf 470}, 486-491 (2011). 




\bibitem{Popescu} N. Linden, S. Popescu, and P. Skrzypczyk, {\it How Small Can Thermal Machines Be? The Smallest Possible Refrigerator}, Phys. Rev. Lett. {\bf 105}, 130401 (2010).

\bibitem{Kosloff} A. Levy and R. Kosloff, {\it Quantum absorption refrigerator}, Phys. Rev. Lett. {\bf 108}, 070604 (2012).

\bibitem{Giovannetti} D. Venturelli, R. Fazio, and V. Giovannetti, {\it Minimal Self-Contained Quantum Refrigeration Machine Based on Four Quantum Dots}, Phys. Rev. Lett. {\bf 110}, 256801 (2013).

\bibitem{Correa} L. A. Correa, J. P. Palao, D. Alonso, G. Adesso, {\it Quantum-enhanced absorption refrigerators}, Sci. Rep. {\bf 4}, 3949 (2014).

\bibitem{Mitchison} M. T. Mitchison, M. Huber, J. Prior, M. P. Woods, and M. B. Plenio, {\it Realising a quantum absorption refrigerator with an atom-cavity system}, Quantum Sci. Technol. {\bf 1}, 015001 (2016).

\bibitem{Hofer} P. P. Hofer, M. Perarnau-Llobet, J. B. Brask, R. Silva, M. Huber, and N. Brunner, {\it Autonomous quantum refrigerator in a circuit QED architecture based on a Josephson junction}, Phys. Rev. B {\bf 94}, 235420 (2016).

\bibitem{Scarani} G. Maslennikov, S. Ding, R. Hablützel, J. Gan, A. Roulet, S. Nimmrichter, J. Dai, V. Scarani, and D. Matsukevich, {\it Quantum absorption refrigerator with trapped ions}, Nat. Commun. {\bf 10}, 202 (2019).

\bibitem{Withney} G. Benenti, G. Casati, K. Saito, and R. S. Whitney, {\it Fundamental aspects of steady-state conversion of heat to work at the nanoscale}, Phys. Rep. {\bf 694}, 1-124 (2017). 



\bibitem{funo} K. Funo, and H. T. Quan,  {\it Path Integral Approach to Quantum Thermodynamics},  Phys. Rev. Lett. {\bf 121}(4), 040602 (2018).


\bibitem{Spekkens} F. G. S. L. Branda\~ao, M. Horodecki, J. Oppenheim,J. M. Renes, and R. W. Spekkens, {\it Resource Theory of Quantum States Out of Thermal Equilibrium}, Phys. Rev. Lett. {\bf 111}, 250404 (2013).

\bibitem{Oppenheim} M. Horodecki, and  J. Oppenheim, {\it Fundamental limitations for quantum and nanoscale thermodynamics}, Nat. Commun. {\bf 4}, 2059 (2013).

\bibitem{Halpern}  N. Y. Halpern, A. J. P. Garner, O. C. O. Dahlsten, and V. Vedral, {\it Introducing one-shot work into fluctuation relations}, New J. Phys. {\bf 17}, 095003 (2015). 

\bibitem{Aberg} J. \AA berg, {\it Fully Quantum Fluctuation Theorems}, Phys. Rev. X {\bf 8}, 011019 (2018).


\bibitem{Kempe} J. Kempe, {\it Quantum random walks: an introductory overview}. Cont. Phys. {\bf 44}(4), 307-327 (2003).

\bibitem{Cirac} F. Verstraete, M. M. Wolf, and J. I. Cirac, {\it Quantum computation and quantum-state engineering driven by dissipation}, Nat. Phys. {\bf 5}, 633-636 (2009).

\bibitem{Davide} G. D. Paparo, V. Dunjko, A. Makmal, M. A. Martin-Delgado, and H. J. Briegel, {\it Quantum Speedup for Active Learning Agents}, Phys. Rev. X {\bf 4}, 031002 (2014).

\bibitem{Jordan} A. Chantasri, M. E. Kimchi-Schwartz, N. Roch, I. Siddiqi, and A. N. Jordan, {\it Quantum Trajectories and Their Statistics for Remotely Entangled Quantum Bits}, Phys. Rev. X {\bf 6}, 041052 (2016).

\bibitem{porras} C. Nation, D. Porras, {\it Quantum Chaotic Fluctuation-Dissipation Theorem: Effective Brownian Motion in Closed Quantum Systems}, arXiv:1811.03028 (2018).


\bibitem{massicoarse}  M. Esposito, {\it Stochastic thermodynamics under coarse graining}, Phys. Rev. E {\bf 85} (4), 041125 (2012).

\bibitem{samurai} N. Shiraishi, and T. Sagawa, {\it  Fluctuation theorem for partially masked nonequilibrium dynamics}, Phys. Rev. E {\bf 91}(1), 012130 (2015).

\end{thebibliography}

\begin{thebibliography}{99}
\bibitem{horowitzs} J. M. Horowitz, and J.M.R. Parrondo, {\it Entropy production along nonequilibrium quantum jump trajectories}, New. J. Phys. {\bf 15}, 085028 (2013),

\bibitem{PRX} G. Manzano, J. M. Horowitz, and J. M. R. Parrondo, {\it Quantum Fluctuation Theorems for Abitrary  Environments: Adiabatic and Non-adiabatic Entropy Production}, Phys. Rev. X {\bf 8}, 031037 (2018).

\bibitem{Neris} I. Neri, \'E. Rold\'an, and F. J\"ulicher, {\it Statistics of Infima and Stopping Times of Entropy Production and Applications to Active Molecular Processes}, Phys. Rev. X {\bf 7}, 011019 (2017). 

\bibitem{Doobs} J. L. Doob, {\it Stochastic Processes} (John Wiley and Sons, Chapman and Hall, New York, 1953); 
J. L. Doob, {\it Measure Theory} (Springer Science and Business Media, Berlin, 1994), Vol. 143.

\bibitem{Williams} D. Williams, {\it Probability with Martingales}, (Cambridge University Press, Cambridge, 1991).

\bibitem{Shamik} \'E. Rold\'an, and S. Gupta, {\it Path-integral formalism for stochastic resetting: Exactly solved examples and shortcuts to confinement}, Phys. Rev. E, {\bf 96}, 022130 (2017).

\end{thebibliography}
\end{document}